# Resonances in the continuum, field-induced nonstationary states, and the state- and property-specific treatment of the many-electron problem

**Cleanthes A. Nicolaides**          February 8, 2016

*Theoretical and Physical Chemistry Institute, National Hellenic Research Foundation, 48 Vasileos Constantinou Avenue, Athens 11635, Greece*

caan@eie.gr

**To be published in Advances in Quantum Chemistry, the 'Löwdin volume', edited by J. Sabin and E. Brändas**

## CONTENTS

--------------------------------------------------------------------------------------------------------

## PREFACE

I start by thanking Erkki Brändas and Jack Sabin for inviting me to contribute to this special volume commemorating the $100^{th}$ birthday of Per-Olov Löwdin. Their initiative adds to previous ones involving conferences and books that have been dedicated to him, all expressing  the respect and admiration that Löwdin inspired throughout his scientific career in his associates and professional colleagues.

Although there are many people who are better qualified to comment on Löwdin's  personality and achievements, I take this opportunity to state briefly my impressions of him.

I met Löwdin only a few times during the 1970s and 1980s, in conferences and in Sanibel symposia, starting with the conference on "The Future of Quantum Chemistry" that was held in Dalseter, Norway, September 1-5, 1976, organized by J.-L. Calais and O. Goscinski to celebrate his $60^{th}$ birthday. Those encounters, (which included a couple of cocktail parties and soccer games where he played goalie), led to a nice rapport, even though I was much younger. They were sufficient to leave me with the best of impressions about his openness, about his interest in assisting young scientists from all over the world, and about his scientific inquisitiveness and aim for mathematical clarity and justification.

Löwdin's scientific and organizational achievements were instrumental in advancing the cause of Quantum Chemistry in Sweden as well as internationally, especially during the 1950s and 1960s. E.g., the Uppsala summer schools and the Sanibel winter symposia became institutions. He is remembered with respect and affection not only for his research papers but also for his exceptional activity which accelerated the recognition of Quantum Chemistry as a distinct scientific discipline with a diverse community of theoretical scientists. As a member of this community, I feel lucky for the opportunity given to me by the Editors to contribute the paper which follows.

# ABSTRACT

The paper summarizes elements of theories and computational methods that we have constructed and applied over the years for the non-perturbative solution of *many-electron problems* (MEPs), in the absence or presence of *strong* external fields, concerning *resonance/nonstationary* states with a variety of electronic structures.

Using brief arguments and comments, I explain how these MEPs are solvable in terms of practical time-independent or time-dependent methods, which are based on single- or multi-state Hermitian or non-Hermitian formulations. The latter result from the *complex-eigenvalue Schrödinger equation* (CESE) theory. The CESE has been derived, for field-free as well as for field-induced resonances, by starting from Fano's 1961 discrete-continuum standing-wave superposition, and by imposing outgoing-wave boundary conditions on the resulting solution. Regularization is effected via the use of complex coordinates for the orbitals of the outgoing electron(s) in each channel. The Hamiltonian coordinates remain real.

The computational framework emphasizes the use of appropriate *forms* of the trial wavefunctions and the choice of function spaces according to the *state- and property-specific* methodology, using either nonrelativistic or relativistic Hamiltonians. In most cases, the bound part of excited wavefunctions is obtained via state-specific '*HF or MCHF plus selected parts of electron correlation*' schemes. This approach was first introduced to the theory of multiply excited and inner-hole autoionizing states in 1972, and its feasibility was demonstrated even in cases of multiply excited negative ion scattering resonances.

For problems of states interacting with strong and/or ultrashort pulses, the *many-electron time-dependent Schrödinger equation* is solved via the *state-specific expansion approach*.

Applications have produced a plethora of numerical data that either compare favorably with measurements or constitute testable predictions of properties of *N-electron* field-free and field-induced *nonstationary* states.

**Keywords:** Resonances, field-induced non-stationary states, many-electron problems, strong fields, time-dependent Schrödinger equation, complex-eigenvalue Schrödinger equation

# 1. QUANTUM CHEMISTRY AND MANY-ELECTRON PROBLEMS IN THE HIGH-LYING PORTIONS OF THE "EXCITATION AXIS"

*For every complex problem there is a solution that is clear, simple and wrong*

H. L. Mencken (1880-1956)

The emergence of Quantum Chemistry (QC) as a distinct discipline has been associated with and driven by the requirement of tackling efficiently and quantitatively the *many-electron problem* (MEP) and its multifarious manifestations in a multitude of electronic structures, of properties and of cases of dynamics of physical/chemical relevance.

Following Wigner's use of the term in his 1934 theory of electrons in metals [1], the MEP also came to be known among quantum chemists as the problem of '*electron correlation*', using as zero-order reference the single-configurational restricted Hartree-Fock (HF) solution and focusing on the accurate calculation of the total energy of the ground state. This definition seems to have been introduced into the literature of QC in 1952 by Taylor and Parr [2], in connection with their discussion on the convergence of the method of '*superposition of configurations*' in the Helium ground state.

Even though the pursuit of the accurate determination of the total energy of the ground state continues to characterize nearly all publications on many-electron methods and algorithms, it should be kept in mind that when it comes to properties and dynamics involving highly excited N-electron states, this is not necessarily rewarding or even feasible.

In **Figure 1** I have sketched two major directions/categories for modern QC which I consider to be distinct in terms of the fundamental theories that govern them and the scientific information which they seek to produce. While the MEP is omnipresent, the physics, the nature of experimental possibilities and the corresponding fundamentals of many-electron Quantum Mechanics in these two "directions" are different.

-------------------------------------------------------------------------------------------

**Figure 1**

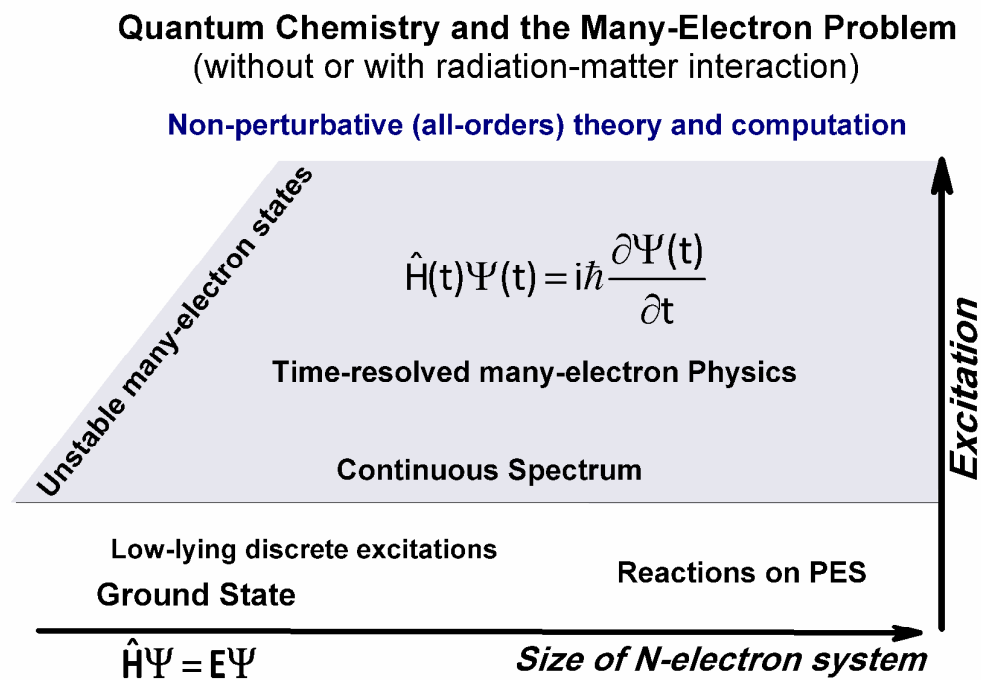

Sketch depicting the two 'directions' of Quantum Chemistry referred to in the introduction. The paper discusses topics in the high-lying portions of the "excitation axis". The corresponding *many-electron time-independent or time-dependent problems* involve field-free or field-induced *nonstationary* electronic states relaxing into the continuous spectrum.

-------------------------------------------------------------------------------------------

The horizontal arrow of **Figure 1** represents the achievements and the continuing extensive activity by practitioners all over the world, which have been pushing research mainly towards larger and larger sizes of atomic and molecular systems in their ground state, with occasional excursions to low-lying excited discrete states. It expresses what has become by far the dominant component of QC in terms of numbers of publications, authors and financial support, namely 'Computational Quantum Chemistry' (CQC). Here, the fundamental equation is Schrödinger's *real-eigenvalue* equation for stationary discrete states,

$$\mathbf{H}\Psi_n = E_n\Psi_n, \qquad <\Psi_n \mid \Psi_m> = \delta_{nm} \qquad (1)$$

When it comes to applications, (countless systems), the utility of CQC extends to materials science and even to biology. Using the relevant approximate solutions $(\tilde{\Psi}_n, \tilde{E}_n)$ and additional theory, electronic/vibrational/rotational spectra (positions and intensities), models of electron distributions, of bonding and of reactivity, other properties, etc., are computed (computable). For some classes of cases, density functional theory is, in principle, also applicable.

As a result of the dominance of CQC, the general perception has been that CQC is essentially the same as QC, with the front moving along the direction where methods and algorithms are improved in order to produce reliable solutions of eq. (1) for electronic ground/low-lying discrete states of molecules with large numbers of electrons. Indeed, using as reference the evolution of methods and of types of applications in CQC, distinct 'ages of QC' have been suggested. However, this identification of QC with CQC can be challenged, on the basis that modern QC is also concerned with novel MEPs in the high-lying portions of the "excitation axis" of **Figure 1**, *near* (below and above) and *far* from the threshold of the continuous spectrum. Arguments supporting this description of QC and MEPs were propounded and documented in [3], see also [4].

The present paper has to do with theories and methods that enable the solution of broad categories of MEPs in the aforementioned realm of the "excitation axis" of **Figure 1**. The topics are stated in the next section. These refer to *resonance* states of finite energy width as well as to discrete *doubly excited states* (DESs) and *Rydberg* series very close to threshold. Furthermore, they involve properties and phenomena induced by *strong* and/or *ultrashort* (few-cycles) electromagnetic fields. In all cases, Hermitian or non-Hermitian nonperturbative (variational) formalisms have been constructed and implemented.

The many-electron quantum mechanics for the correct formal and computational treatment of the plethora of situations that these topics contain, is much more sophisticated and demanding than that which applies to ground or low-lying discrete states. Now, given the level of complexity of the general formalisms which have to account for the continuous spectrum of scattering states in conjunction with a multitude of open-shell electronic structures, the difficulties and challenges of the MEP not only persist but are accentuated considerably. For example, consider the *nonstationary* states that are created by the interaction, $V(t)$, of atomic/molecular ground or excited states with *strong/ultrashort* electromagnetic pulses which drive the system into the continuous spectrum. In order to understand quantitatively such phenomena, one must pursue the *non-perturbative* solution of the *many-electron time-dependent Schrödinger equation* (METDSE),

$$H(t)\Psi(t) = i\hbar \frac{\partial \Psi(t)}{\partial t} \qquad H(t) = \mathbf{H}_{A(M)} + V(t) \qquad (2)$$

Now, the MEP becomes time-dependent, and issues of the interplay between dynamics and electron correlations are resolved on the time axis [4].

I close this introduction by noting that, because of the nature and scope of the paper, the presentation is necessarily condensed. Furthermore, the reference list is inevitably dominated by publications of the author, without or with co-authors. This is necessary in order to support the brief arguments and to guide the reader accordingly. The work by other researchers which is cited, is directly connected to the statements and comments.

## 2. TOPICS

The paper comments briefly on the topics I, II and III listed below in the chronological order in which they were introduced in the literature, and on the formalisms and computational methods which we have proposed and applied towards the solution of the corresponding MEPs for states, properties and processes. Extensive recent discussions, including characteristic results of calculations that we have published on prototypical systems and problems, are presented in [3-7].

*I)* Theory and calculation of the properties of N-electron *field-free resonance states* ('autoionizing', or 'Auger', or 'negative-ion compound' states, etc.), of the '*Feshbach*' or the '*shape*' type. In principle, these states can be (and have been) created and measured in excitation/collision processes of various types. In addition, this topic includes the *multistate* theory and calculation of spectra just below and above threshold, which, however, cannot be discussed here, due to the constraints on the length of the article (see section 5.1).

*II)*  Theory and calculation  of the *non-perturbative* response (energy shifts and ionization rates) of ground and excited discrete states, as well as of resonance states, to *strong static or periodic* electric or magnetic fields.

*III)*  Theory and calculation of the *time-resolved*  physics of N-electron nonstationary wavepackets, determined based on  the *non-perturbative* solution  of the METDSE, for time-independent Hamiltonians, (as is the case of autoionization or of decay due to perturbation by a static external field), and, principally, for time-dependent Hamiltonians describing the perturbation of N-electron ground or excited states by *strong  and/or ultrashort* electromagnetic pulses.

## 3. STATE- AND PROPERTY-SPECIFIC QUANTUM CHEMISTRY

The goal in the design and execution of our treatments of topics I, II and III has been, on the one hand the formulation of general as well as computationally tractable  theoretical frameworks, and, on the other hand, the consistent and reliable tackling of issues of electronic structure, of electron correlation, and of the continuous spectrum. The choices and optimization of basis sets (analytic as well as numerical orbitals), and the calculation of the wavefunctions and of the matrix elements that are dictated by theory, is done according to the '*state- and property-specific*' (SPS) quantum chemistry [7], also [3-6].

The basic argument of the SPS theory has been that, in order to render the solution of a variety of N-electron time-independent or time-dependent problems economic or even feasible, of primary importance is the choice of  *forms* of trial wavefunctions and the optimization of the function spaces that are relevant to the state(s) and to the property of interest. This means that, in many cases, it is not necessary to pursue a very accurate number for the total energy. Instead, the degree of accuracy which is expected from the approximate solution of the pertinent Schrödinger equation(s) with respect to the total energy varies depending on the state(s) and on the problem.

For example, as proposed and demonstrated computationally in the 1970s [7], the calculation of radiative and radiationless transition probabilities can very often be accomplished reliably by analyzing and understanding directly the off-diagonal matrix elements, the state-specific orbitals and mixing coefficients, and the possible N-electron overlaps, rather than by first aiming at a very accurate calculation of the total energies and wavefunctions.

Critical to the success of this endeavor, especially when it comes to excited states, is the definition and accurate calculation of a zero-order '*Fermi-sea*' wavefunction, normally obtained via the solution of the state-specific multi-configuration Hartree-Fock (MCHF) equations [7]. The concept of the *Fermi-sea* self-consistent orbitals and corresponding symmetry-adapted configurations was introduced in the 1970s, with prototypical applications to  simple excited states of atoms [7].

In other words, the focus in the treatment of the various MEPs is *not* merely on the construction of formalism by manipulating various operator/matrix expressions, expecting that these can, somehow, be computed accurately under the *assumption* that there exists a convenient, (nearly) complete set (Hilbert space) on which they are defined. Instead, having established a formally sound framework, the emphasis is first on how the total wavefunction of each state can be discerned in terms of the significance of its main components, and of corresponding basis sets, regarding the problem and the property under investigation. This means that one focuses mainly on mixing coefficients of symmetry-adapted configurations and on the numerical accuracy of appropriate zero-order and correlation orbitals.

Beyond the HF or the MCHF zero order approximation, the contributions of symmetry-adapted components describing electron correlation are computed via diagonalization of Hamiltonian matrices that are normally small. In practice, this leads to reliable *state-specific* wavefunctions that are generally compact, numerically accurate and transparent. Consequently, they are suitable for  applications, and, often, for transferring information from one system to another.

A case in point is the calculation of wavefunctions and properties of resonance states. Starting in 1972 [8], a series of publications have demonstrated the critical importance of properly optimized (according to theory) function spaces, especially when computer facilities are very limited [5-7].

It is worth pointing out that from the choice of the *Fermi-sea* wavefunction depends the further separation of the remaining electron correlations into '*non-dynamical*' and '*dynamical*'. These concepts, which are often discussed and used in the recent literature of molecular electronic structure calculations, were introduced and first applied to low-lying discrete states of small atoms by Sinanoğlu and co-workers in the 1960s, who used the more restricted concept of the '*Hartree-Fock sea*' of spin-orbitals, which had been based on the filling of single hydrogenic or 'minimal basis' shells [20]. The discussion in [7] on these concepts includes as an example the analysis of a rather difficult bonding situation, namely that which characterizes the weakly bound $Be_2$ molecule.

The essentials of the SPS theory are discussed in [5-7]. Many of its applications have been concerned with atomic properties and phenomena, with emphasis on topics I, II and III. Aspects of its methods have also been demonstrated in problems with diatomics and polyatomics [5-7]. For the latter, there are open and challenging issues for theory as well as for computational methods.

## 4. BACKGROUNDS

The following paragraphs recall the state of affairs with theory and calculations at the time of the introduction of the analyses, formalisms and SPS methods for the quantitative treatment of MEPs in the topics I, II and III. Any other approaches which may have been published in subsequent and recent years are not in the scope of this review, regardless of the degree of their overlap with the main ideas and results of the many-electron approaches discussed here.

For topic I, the reference period is up to about the end of 1971, which is when the *decaying-state, state-specific theory* and results were completed, and published in [8]. For topic II, the reference period is up to 1988, which is when we began publishing our first papers on the stationary '*many-electron, many-photon theory*' [9-14]. For topic III, the reference period is up to 1994, which is when our first papers on the '*state-specific expansion approach*' (SSEA) to the solution of the METDSE were published [15-17].

### 4.1. Topic I: Field-free resonance states and the MEP

As one can verify from a perusal of the QC literature up to 1971, the achievements of quantum chemists during the decades of the 1950s and 1960s on the MEP were not in this area. Instead, the focus was on the development and analysis of methods that solve eq. (1) for $(\tilde{\Psi}_n, \tilde{E}_n)$ of ground and low-lying discrete states, based on the scheme '*HF plus electron correlation energy*'. Eg., see [18], whose 13 review articles embody the bulk of proposals, trends and progress, up until the end of the 1960s, towards the understanding of the MEP associated with the solution of eq. (1). (While a PhD candidate at Yale, (1969-1971), I familiarized myself with the substance and the details of most of those articles. The experience which I acquired from the study and implementation of Roothaan's analytic HF method [19] and of Sinanoğlu's theory of electron correlation in ground and low-lying discrete states [20], was instrumental in my taking the first steps of the *state-specific* treatment of inner-hole and of multiply excited states (even of negative ions), in the decaying-state framework of [8] – see present section 5).

At the same time, (1950s and, mostly, 1960s), the theme of "resonances" in atomic and molecular physics was emerging both on the experimental and the theoretical front. (E.g., see the reviews [21,22]). The bulk of formalisms and methods followed the path of scattering theory, to the extent that they could handle few problems of low-lying resonance states in two- and three-electron systems, [21-26]. In addition, bound-state-type methods, generally known as the *diagonalization/stabilization method*' (DSM), had also been introduced and studied, [27-31]. The common characteristic of those approaches was that, regardless of the level of rigor of the formalism, the implementation was done by using a single basis set, either for just the bound wavefunctions or for both the bound and the scattering components (as is the case of the DSM).

Out of the many possible examples of work from that period, of interest to the reader may be the CI calculations and conclusions of Hylleraas [27] on a couple of low-lying DESs of $H^-$. This 1950 paper is probably the origin of publications using some version of DSM. Hylleraas [27] dealt with the problem by using only discrete basis functions and searching for the "correct" roots of the diagonalized energy matrix. The uncertainties, the concerns and the approximations which can be found in that paper, by the man who 20 years earlier had produced the remarkably accurate calculations on the ground state of Helium and its isoelectronic ions, are testimony to the subtleties and difficulties (in addition to lack of computer power) with which the few people in the 1950s-60s who had tackled the problem of actual calculation of resonance states were faced.

In summary, as explained and documented in [8], and also recently in [5], the theories and/or methods of calculation of wavefunctions, energies and widths (total and partial) of resonance/autoionizing states which had been published by 1971, were characterized either by formal incompleteness or by serious practical difficulties concerning their applicability to N-electron structures, (with electrons excited either from the valence or from the inner shells), while incorporating the corresponding electron correlations.

Given this background, the approach which was introduced in [8] and since then improved and expanded [5], showed how it is possible to incorporate advanced polyelectronic methods for arbitrary electronic structures and for electron correlations in resonance (autoionizing) states of N-electron atoms and molecules (N > 2), into theory that draws from the general theory of scattering and accounts for the presence of the continuous spectrum.

## 4.2.  Topics II and III: Ground or excited states perturbed by *strong* static or periodic or pulsed fields, and the non-perturbative solution of the MEP

In topics II and III, $V$ of eq. (2) may or may not be time-dependent. In principle, the theory for the quantitative treatment of the perturbation of ground or excited states by *strong* fields must be either non-perturbative with respect to the total Hamiltonian, $\mathbf{H}_{A(M)} + V$, or capable of summing accurately the time-independent or time-dependent series of perturbation theory to all orders. Hence, it necessarily engages, directly or indirectly, a multitude of states from both the discrete and the continuous spectrum, whose degree of mixing and the concomitant effects on observables depend on the spectrum of $\mathbf{H}_{A(M)}$ and on the parameters of the external field. The main result of this field-induced mixing is that the system is driven into the continuous spectrum, i.e., it releases one or more electrons.

For static and for periodic fields, the METDSE is reducible to stationary forms. The energies of the state(s) of interest not only are shifted but are also broadened because of the mixing with components of the continuous spectrum. Thus, discrete states obeying eq. (1) turn into states with characteristics of *resonances*, where the energy shifts and widths are field-dependent. At the same time, new features in the spectrum may appear, as a function of, say, the intensity, for fixed frequency.

For ultrashort and/or for strong interactions caused by pulsed fields of few periods, the physics is necessarily time-dependent, and the changes in the old levels (and the appearance of new ones) occur at each moment during the interaction. Now, an originally stationary state becomes a field-dependent nonstationary wavepacket, $\Psi(t)$, obeying eq. (2). Projection of $\Psi(t)$ onto stationary states of the discrete and/or the continuous spectrum after the end of the pulsed interaction gives the occupation probabilities of these states.

The phenomena of electron emission corresponding to the above situations are known as *tunneling* or *(multi)photoionization*. Associated with them are shifts of the energies of the field-free states, (for weak fields, these yield static and dynamic (hyper)polarizabilities), and time-independent rates or time-dependent probabilities of transition into the continuous spectrum.

Because of the severe complications that result from the addition of strong $V$ to $\mathbf{H}_{A(M)}$, until the late 1980s no practical theory and computed results existed showing how to perform non-perturbative, (or, all-orders in perturbation theory), calculations for field-induced resonances and/or nonstationary involving arbitrary closed- or open-shell many-electron states, including the effects of electron correlation.

In the context of the foregoing comments, two theoretical advances from the 1970s and 1980s must be recalled. These concern the non-perturbative theory and calculation of properties in both topics. Their introduction was done at a time when it had become clear that the experimental progress with intense laser at various wavelengths would require the development and application of non-perturbative theories and methods.

### 4.2.1 Topic II: Complex-energy variational methods for static and periodic strong fields

Following the 1971 mathematical results of Aquilar, Balslev and Combes, as clarified by Simon, regarding the spectrum of the 'dilatationally analytic' Coulomb Hamiltonian upon 'rotation' of its coordinates which are allowed to be complex, $r \rightarrow re^{i\theta}$ [32], Doolen [33] proposed the variational *complex coordinate rotation* (CCR) procedure, whose capacity he demonstrated by computing 'a resonant eigenvalue in the $e^-H$ singlet $s$ wave' using a basis set of Hylleraas functions.

In principle, this 'rotation' reveals the full spectrum of discrete, scattering and resonance states. Reinhardt and co-workers [34-37] found via computation on Hydrogen that the CCR method also works for the electric dipole operator of strong dc and ac fields, which is not dilatationally analytic. As regards 'Hydrogen in ac fields', Chu and Reinhardt [36] invoked the work of Shirley [38] on the reduction of problems with periodic time-dependent Hamiltonians to stationary problems in the 'semiclassical' approximation, ('Floquet methods'), and demonstrated the physical relevance of complex eigenvalues of the electric dipole 'Floquet Hamiltonian', $H_F(\theta)$, by computing widths of field-induced resonances representing multiphoton ionization rates.

The approaches and findings by Reinhardt and coworkers during the 1970s constituted a significant step forward in the theory of atom-strong field interactions, since they were based on a variational method which takes into account the continuous spectrum. However, when it comes to the spectra of nearly all of atoms and molecules, the use of the CCR method is unrealistic, since the recipe demands a large number of diagonalizations of $H(\theta)$ on a large basis set and the search for the properties of all roots as a function of $\theta$ and of other variations in the basis.

The practical limitations of the CCR method regarding the solution of MEPs for arbitrary electronic structures and for the calculation of partial widths, for complex Hamiltonians without or with an external field, was pointed out in [39,40,5]. One of the main reasons is that the use of a common basis set in terms of which the Hamiltonian matrix is constructed, cannot represent economically and accurately the function spaces which characterize on resonance the '*localized*' and the '*asymptotic*' parts of the correct solution [5]. In other words, just as in the ordinary case of N-electron ground states the brute-force diagonalization of **H** on a big basis set runs quickly into trouble, the requirements associated with the CCR method could not (cannot) render it suitable for the treatment of MEPs for arbitrary electronic structures. In fact, when the N-electron Hamiltonian includes an external ac field, the practical difficulty of the CCR task increases beyond reasonable levels, since, in addition, huge 'Floquet blocks' would have to be constructed and tested, via repeated diagonalizations, seeking convergence of many roots.

The aforementioned limiting difficulties regarding the solution of the MEP cease to exist in the polyelectronic theory which was proposed in the 1970s as a natural extension of the work that had started in [8]. This is the *state-specific 'complex eigenvalue Schrödinger equation'* (CESE) *approach*, [5], features of which are explained in section 5.

The basis of the CESE method is the theory-guided emphasis on recognizing the two-part *form* of the resonance eigenfunction and on pursuing the separate optimization of the corresponding function spaces. Now, the coordinates of the Hamiltonian are real, and so are the coordinates of the correlated wavefunction for the *localized* part, $\Psi_0$. Only the *asymptotic* part of the trial wavefunction in the appropriate channel is made complex in order to regularize the solution on resonance. Its extension to problems with external static or ac fields, named the *many-electron, many-photon theory*, uses similar arguments and methods with a time-independent Hamiltonian that includes the interaction. (The matrix elements are, of course, different) [9-14,5]. Computational comparisons of the CCR and the CESE methods which demonstrate the efficiency of the CESE method even in simple cases, where the CCR method is applicable, were published in [41,42].

### 4.2.2   Topic III: 'Grid' methods

As the science and technology of the production and spectroscopic use of electromagnetic pulses that are intense and/or ultrashort is advancing, the horizon of the scientific area of light-matter interactions is widened significantly, especially with respect to prospects of obtaining new information on multi-electron systems. The comprehensive quantitative understanding, via theory, of corresponding effects (mainly multiphoton and pump-probe processes), presupposes the capacity of obtaining and using the $\Psi(t)$ of eq. (2), describing the nonstationary state of interest.

In the 1980s, Kulander [43,44] implemented the 'grid method' for integrating numerically the METDSE, via two approximate schemes which invoke the independent-electron model. The first was the time-dependent HF approximation, which was applied to He [43]. The second was the '*single active electron*' approximation, where the time-dependent wavefunction of only one electron moving in an average local potential is computed. It was applied to the calculation of the time-dependent multiphoton ionization of Xenon [44]. The calculations and analysis of Kulander provided for the first time useful information on specific multiphoton processes in multielectron atoms. Yet, both of these schemes were  limited to closed-shell, single determinantal states, (e.g., noble gas atoms). They are "one shot" approaches which ignore the MEP even for closed-shell states, since they do not take into account the electron-correlation and interchannel-coupling components of electronic structures and of transition amplitudes.

In 1994, the Athens team published a polyelectronic theory, (the SSEA), with numerical results, which shows how the METDSE can be solved systematically for arbitrary electronic structures [15,16,6]. The SSEA is not a 'grid' method. (See section 5). The initial applications involved two proof-of-principle cases (apart from Hydrogen). The first was to the correlated wavefunction of the $Li^-$ $1s^2 2s^2$ ground state, with two ionization coupled channels, ($1s^2 2s\ ^2S$ , $1s^2 2p\ ^2P^o$) [15]. The second was to $He$ $1s2s\ ^1S$ , which is the prototypical open-shell, metastable excited state having a lower state of the same symmetry [16]. Problems of field-induced time-dependent dynamics in diatomic molecules were also solved at that time [6].

Furthermore, by implementing the SSEA, it also became possible for the first time to compute from first principles the *non-exponential decay* of real, multiparticle unstable states, where the matrix elements involve only the time-independent $\mathbf{H}_{A(M)}$ and appropriately computed N-electron wavefunctions [17].

## 5. OVERVIEW AND ELEMENTS OF THE SPS THEORY ON TOPICS I, II AND III

The contents of the previous sections have much to do with the "why" our SPS work on topics I, II and III was undertaken, in view of the desideratum to treat the corresponding MEPs quantitatively within rigorous theoretical frameworks. In this last section, I will outline only a few elements of "what" has been done and "how" it was (is) done. The extent of the presentation is very limited, focusing mostly on Topic I, because of space restrictions. Apart from the original publications, much of the important information can be found in the recent reviews [3-7].

### 5.1. Topic I

Hermitian as well as non-Hermitian (CESE theory) formulations have been constructed and applied for the prediction of energies, widths (partial and total) and oscillator strengths of resonances due to singly or multiply excited states and to inner-hole (Auger) states, of neutrals and of positive and negative ions, mainly of atoms, but also for characteristic cases of negative-ion resonances and of *diabatic* states in diatomics such as $He_2^-$ and $He_2^+$ [5].

In addition, certain of the features and results of the SPS approach can also be useful for treating problems in nuclear and in solid state physics. Indeed, such applications have already taken place. For example, in [45] we reported energies and analysis of the wavefunctions for inner-hole states of clusters simulating the real metal. In [46], the implementation of the CESE theory produced tunneling rates in multi-barrier solid state structures relevant to properties of devices. Recently, [47], the results from the time-dependent decaying-state model of [39] proved relevant to an analysis of nuclear decays, by answering the challenge [48] to the long-honored method of radioactive carbon dating which has been based on the assumption of exponential decay of the $^{14}_{6}C$ isotope.

The problems for which quantitative solutions have been produced include not only cases of isolated states but also complex spectra due to *multistate* interactions in multichannel continua, even in cases where the binding is weak and the widths are very small. I give two examples:

1)  The CESE work and the analysis of results reported in [49-51] achieved the first definitive resolution and interpretation of the spectrum of the complex eigenvalues representing the low- as well as the high-lying resonance states of $H^-$, (up to the $n = 5$ hydrogen threshold) for which the zero-order model had been proposed about 40 years earlier ('*dipole resonances*') [52]. It is worth stressing that this is the first real system where essentially the complete resonance spectrum has been resolved. Among other things, this achievement required the appropriate use of (very) diffuse functions that went out to 8000 a.u., a fact which, in the context of the CESE theory, allowed the calculation of widths down to $10^{-9}$ au..

2)  In a series of papers, [53-57] and section 5.2 of [6], a CI, K-matrix formalism which unifies, in a computationally tractable manner, the treatment of *multi-state* interactions in the discrete and the continuous spectra, including multiply excited states, was proposed and applied, with emphasis on the possibility of calculating wavefunctions and spectroscopic data just below and just above the fragmentation thresholds.

For example, [54] reported the cross-sections for the simultaneous photoionization and photoexcitation of $He$, around and at the $He^+$ $n = 2$ threshold, as well as the positions and autoionization widths of three channels of $^1P^o$ Rydberg series of resonances.

Another example can be found in [55,56], where, using the Breit-Pauli Hamiltonian and MCHF-based CI wavefunctions, we made the first quantitative predictions of the wavefunction character, the quantum defects, the oscillator strengths and the fine structure of the $A\ell$ $KL3s^2nd$ $^2D$ Rydberg series perturbed by the $KL3s3pnp(\varepsilon p)$ $^2D$ channel and, especially, the valence '$KL3s3p^2$' $^2D$ correlated state in the presence of the $KL3s^2\varepsilon d$ continuum.

Regarding the above multielectron, multistate problem, it is worth noting the following: The valence configuration $KL3s3p^2$ was found to be spread over the lower portion of the discrete spectrum and not to be located just above the $A\ell^+$ threshold, as concluded by Taylor, Bauschlicher and Langhoff [58] based on the results of standard 'full CI' calculations. The significance of this disagreement goes beyond the fact that two methods gave different results. Indeed, it has to do with the fundamentals of quantum mechanical calculations regarding strongly perturbed spectra, and with the degree of validity of standard methods of CQC with common basis sets, such as the one applied in [58], which ignore the particularities and details of the Rydberg series and of the continuous spectrum. Our predictions were later confirmed experimentally in [59].

The CI-K-matrix SPS theory of [53-57] was developed by drawing from the quantum defect theories of Seaton [60], and of Fano [61] and Greene et al [62]. However, it differs from them in essential ways, especially as regards the calculation of bound and scattering wavefunctions and K-matrices for N-electron systems. One of them is that it is formulated and implemented without having to refer to irregular Coulomb functions, as do the formalisms in [60-62]. Such functions are not used in the CI-type calculations which are necessary when it comes to the solution of problems of spectra involving the mixing of N-electron basis wavefunctions.

### 5.1.1. Decaying-state theory as framework for the description of N-electron resonance states

Paper [8] is the origin of this research program, having as immediate sequels [63,39,40]. It is characterized by two main proposals, which are outlined in the following paragraphs. The first is the general theoretical framework for the formal and practical analysis, understanding and calculation of resonance (autoionizing) states. The second is the incorporation into this framework of electronic structure methods based on the scheme of "*state-specific, open-shell, multideterminantal Hartree-Fock plus electron correlation*", for the description of the localized part, $\Psi_0$, of such states, even though the Ritz minimum-energy principle is not applicable.

Following a critical discussion on the concepts and on the then existing methods [8], it was concluded that the time-dependent concept and formalism of *decaying states* can provide the framework which justifies a unified treatment for arbitrary electronic structures (with electron correlation) whose energies are in the continuous spectrum. The formalism produces the result of the complex-energy simple pole to which a resonance state corresponds, $z_0 = E_0 + \Delta - \dfrac{i}{2}\Gamma$, in terms of well-defined matrix elements for the energy of the initially ($t = 0$) localized wavepacket, $\Psi_0$, $E_0 = <\Psi_0 | \mathbf{H} | \Psi_0>$, the energy shift, $\Delta$, and the width, $\Gamma$, in analogy with the resonance scattering theory of Feshbach [64,65]. Using this formalism, physically relevant results have been obtained on the energy- as well as on the time-axis, for a model in which the '*self-energy*' of the autoionizing state, $A(z)$, is approximated by $A(z) \approx A(E_0)$. For example, the analysis of the time-dependence of the isolated decaying states presented in [39] has led to conclusions regarding the possible observation of non-exponential decay for unstable states very close to threshold, as well the foundations of the theory and the possible manifestation of the connection of non-exponential decay to the issue of time-asymmetry [5,47,66].

The formal description of the decaying state in [8,39] was adapted from that of the theory of radiation damping of Heitler [67] and of Schönberg [68], and from the presentation of Goldberger and Watson [69]. In spite of the fact that the formal approach to the decay due to atom (molecule)-field interaction and to autoionizing states is nearly the same, (e.g., the choices of the density of states differ), there is a critical difference, conceptually and practically, which adds difficulty to the case of autoionization. This is the following: Whereas in the problem of atom-field interaction the initial and final states are defined precisely as the eigenfunctions of the unperturbed, field-free Hamiltonian, (i.e., these are stationary states of eq. (1)), in the case of N-electron resonances no such separation is offered, since, upon preparation of the initial state, the interaction causing the decay is intrinsic to the system. In other words, the initially prepared ($\Psi_0, E_0$), is an *N-electron wavepacket* and not a stationary state of the exact operator $\mathbf{H}$, and the question is how to construct it and compute it as efficiently and accurately as possible.

Finally, it is worth pointing out that in the context of the decaying-state theory, the quantum motion is obtained to all orders of time-dependent perturbation theory, since the survival amplitude for decay is defined by the Fourier transform, over the energy spectrum, of $<\Psi_0 | 1 / (z - \mathbf{H}) | \Psi_0>$. Provided that $\Psi_0$ is computable, the theory is formally consistent, and avoids the 'mathematical' questions of convergence of the time-dependent perturbation series for autoionization which became the object of study in [ 37,70], around the time the work of [8] was published. In fact, this is done without resorting to the unphysical and computationally naïve mathematical assumption of a zero-order Hamiltonian which is hydrogenic, as was done in [37,70].

### 5.1.2. Essentials of the SPS approach

The introductory discussion of [8] first pointed to some of the peculiarities and difficulties that characterize resonance states of N-electron atoms and molecules. E.g., I quote: '....*In addition, the useful property of one-to-one correspondence between reality and configurational assignment may in some cases be lost as one moves up in energy and the density of states increases. Also, from the mathematical point of view, they do not have the useful property of square-integrability, having outgoing radiation boundary conditions. Thus, they do not form a complete orthonormal set and variational or perturbation theories dealing with such states must essentially be non-Hermitian in character*'. (Page 2079 of [8]).

In accordance with the last statement, the theory of [8] starts by *asserting* a CESE which the many-electron atom is expected to obey in a resonance state:

$$\mathbf{H}\Psi = W\Psi\,, \qquad W = E_r - i\Gamma/2 \qquad \text{(eq. 1 of [8])} \qquad (3)$$

$$\Psi_{r\to\infty} \sim e^{ik_m r} \qquad (4)$$

$k_m$ is the complex momentum for decay channel $m$ (eq. 2 of [8]).

The appearance of a complex eigenvalue in a Schrödinger equation when $\mathbf{H}$ is formally Hermitian need not cause concern as regards the principles of quantum mechanics. It is explained by the fact that the asymptotic boundary condition on the solution of eq. (3) is not that of square-integrability, (Hilbert space), which is the quantum mechanical condition for the function space on which the real-eigenvalue eq. (1) for discrete states is valid. Instead, assuming one outgoing electron in a particular channel, the eigenfunction satisfies the Sommerfeld outgoing radiation boundary condition, e.g. [71], with a complex energy (complex momentum), which Siegert published in the context of scattering theory [72]. Siegert's result emerged as a complex pole of a model of an s-wave scattered by a short-range potential (nuclear physics), which, however, has limited relation to dynamics involving N-electron systems with Coulomb interactions.

The association of resonance states with a CESE, does not gainsay the fact that, following Fano's Hermitian formalism [73], they can also be treated in terms of real functions and real energies. Below, I will explain how the non-Hermitian eq. (3) can be obtained rigorously, together with a two-part form of the complex eigenfunction, by using as the starting point Fano's theory, which does not depend on models (such as that of Siegert) and is not restricted by the type of the potential or by the types of excitation and electronic structures.

The next initial step in the theory of [8] was the projection of $\Psi$ onto $\Psi_0$ defined above. This means that what is left out is the *asymptotic* component, $X_{as}$, of the resonance eigenfunction, $\Psi_r$, which carries the information of the complex eigenvalue, according to eqs. (3,4). Therefore, the argument can be made that, whether in wavefunction or in operator-matrix representation, the *form* of the N-electron $\Psi_r$ that best describes the physics *on resonance*, is,

$$\Psi_r = \alpha\Psi_0 + X_{as} \tag{5}$$

The square-integrable $\Psi_0$ results from '*dynamic localization*' and contains most of the information pertaining to the character of the system, including those coming from the continuum, before the residual interaction causes the mixing with other states. Its accurate knowledge is critical for the reliability of the overall calculation of the energy and, especially, of the total and partial widths and of other matrix elements. Depending on the problem, the coefficient $\alpha$ and the 'asymptotic' part, $X_{as}$, are functions of the energy, real or complex. The symbol $X_{as}$ represents both the terms of the wavefunction expansion and their coefficients. The additional mixings with other states can be included in terms of multistate formalisms [49-51, 53-57].

The form (5) is fundamental. The SPS formalisms are rendered computationally tractable by focusing on the theory-guided appropriate choices of function spaces for $\Psi_0$ and for $X_{as}$. These are different and are optimized separately, while keeping them as *state-specific* as possible.

The label '*state-specific*' for the approach introduced in [8], refers mainly to the calculation of the correlated ($\Psi_0, E_0$). By bypassing the insurmountable obstacle which characterizes any method that depends on the brute-force diagonalization of the Hamiltonian (real or complex) on a single basis, it introduced and demonstrated the feasibility for obtaining the wavefunctions of such states (any open-shell electronic structure) in terms of the scheme '*HF plus localized and asymptotic electron correlation*', and, soon afterwards, in terms of the scheme '*MCHF plus remaining localized and asymptotic electron correlation*' [5,40].

In most cases, the decay comes from pair correlations, and these have both localized and asymptotic ('*hole-filling*') components. For '*shape*' resonances, the open channel is represented by a single-electron correlation function [5].

In this context, crucial step is the direct solution, (no root searching), of the *state-specific* HF or MCHF equations for the electronic configuration(s) of interest, expected to represent, *in zero-order*, $\Psi_0$. If properly converged, the square-integrable solution does so into a local minimum, satisfying the virial theorem inside the continuous spectrum, and, if necessary, is obtained subject to appropriate orbital orthogonality constraints [5,8,74,75].

Thus, by breaking down the MEP in terms of a zero-order HF or MCHF wavefunction and remaining correlation, it has been argued and demonstrated that is possible to gauge the calculation according to fundamental elements of polyelectronic theory. For example, in the abstract of [8] one reads: *This method requires projection onto known one-electron zeroth-order functions and it thus overcomes the difficulties of the well-known P, Q methods which require projection onto exact wave functions.*

That the HF and MCHF equations can be solved reliably for a variety of highly excited open-shell electronic structures even for weakly bound systems, such as $He^-$ '$2s^2 2p$' $^2P^o$ and '$2s2p^2$' $^2D$, was first demonstrated in [8], at a time when this was far from obvious and computational possibilities of trial and error were very limited. I stress that such HF or MCHF calculations require proper care and numerical accuracy in order to ascertain *localization*, on which the remaining localized correlation is built variationally in order to produce the final $\Psi_0$ of eq. 5 [5, 75].

Of course, for the much simpler cases of relatively isolated configurations with strong attractive potentials, e.g., $1s2s^2 2p^n$ ($n = 1,2..$) Auger states, construction of the corresponding HF equations and convergence of their solution are straight forward, either nonrelativistically or relativistically. This fact has been utilized in a series of many-electron calculations of one-electron energies, of Auger energies and widths, and of radiative decay (fluorescence yield), where electron correlation and relativistic effects are accounted for. E.g. [45, 78, 80, 89, 90]. Characteristic example, with comparisons with other theoretical methods and with experiment, is the study of the wavefunction, electron correlation, energy position and width of the $Be^+ 1s2s^2$ $^2S$ Auger state [78].

In more recent years, the calculation of inner-hole single- or multi-configurational HF or Dirac-Fock atomic wavefunctions and energies is used routinely. Nevertheless, of significance are specific correlation effects beyond such approximations, e.g., [90].

I point out that, in computing an optimal $\Psi_0$, contributions from electronic configurations representing portions of the continuum can also be included without destroying its localization. In the context of the SPS theory, this is often done at the zero-order MCHF level, by including into the equantions '*open-channel like*' configurations [5,75]. It was first done approximately in [8], via diagonalization of small matrices where nearly-degenerate configurations representing *non-dynamical* correlations were included.

$X_{as}$ contains bound (N-1)-electron core wavefunctions coupled to open-channel, term-dependent orbitals representing the outgoing electron in each channel. It may or may not be optimized in the presence of $\Psi_0$, depending on the problem and on the methodology that is implemented. The coefficients of $\Psi_0$ and $X_{as}$ in the resonance state are determined by their mixing in a final step of the calculation. Depending on the problem, either $\Psi_0$ alone or both $\Psi_0$ and $X_{as}$ are used for the calculation of matrix elements and of corresponding properties. For total and for partial widths, the state-specific nature of the overall calculation allows, among other things, quantitative analyses that reveal the dominant effects of strongly mixing zero-order configurations, of overlaps due to the nonorthonormality of orbitals, and of quantum destructive and constructive interference and cancellation effects, without and with interchannel coupling.

Calculations of energies and of partial and total widths of a variety of low- and of high-lying resonances have been carried out on the real energy axis and in the complex energy plane, for nonrelativistic and relativistic autoionization of atoms, e.g., [5, 40 , 40-51, 75-86], as well as for predissociation of diatomics, e.g., [87].

I note that, for some problems it suffices to compute and use only $(\Psi_0, E_0)$, if it is deemed that it carries to a good approximation the information which is needed for the calculation of the quantity of interest. For example, this is the case of the total energy when the energy shift due to the interaction with the part of the continuum represented by $X_{as}$ is expected to be relatively unimportant for the problem of interest. This understanding has allowed systematic work on various properties such as,

- Analysis of wafunctions and regularities of energy spectra of doubly, triply or even quadruply excited states, eg., [88].

- Analysis and computation of one-electron binding and Auger energies, e.g., [45,89,90].

- Calculation of potential energy curves of 'Feshbach' or 'shape' resonances of negative ions of diatomics [91], etc.

### 5.1.3. The complex-eigenvalue Schrödinger equation (CESE)

I now return to the CESE, eq. (3). Its adoption leads to an issue which is important for the understanding and the calculation of resonance states and whose resolution was presented in 1981 [40]. I explain:

According to quantum mechanics, the Hermitian **H** has a complete set of stationary states, consisting of the union of the discrete states of eq. (1) and the scattering states of the continuous spectrum. The latter satisfy the real-energy Schrödinger equation, but not as solutions of an eigenvalue equation. Instead, they have scattering boundary conditions obeying Dirac normalization:

$$\mathbf{H}\Psi(E) = E\Psi(E), \qquad <\Psi(E)\,|\,\Psi(E')> = \ \delta(E - E') \qquad (6)$$

In his classic 1961 paper, Fano [73] presented a Hermitian formalism of CI in the continuum, where the energy-dependent phenomenology of resonances in terms of matrix elements emerges rigorously on the real-energy axis. The questions of the MEP and of ab initio calculation were avoided in [73], since the basis N-electron wavefunctions and their energies were assumed known. The derivations in [73] observed the conditions of eq. (6).The complex eigenvalues for resonances that had occasionally been discussed in the scattering-theory literature of nuclear physics, e.g., [92], were not even mentioned by Fano.

So, a question which has to do with the formal "nature" of resonance states as well as with criteria that must be considered when attempting their calculation in N-electron systems, is the following: Given the rigor and physical relevance of Fano's formalism which is done on the real-energy axis, and given that the complete spectrum consists of states that satisfy eqs. (1) and (6), where are the complex eigenvalues of **H** and the corresponding resonance wavefunctions "hidden", and how can they be revealed and calculated? (I recall that many-body systems cannot be subjected to the "easy", even analytic, theoretical descriptions that physicists often apply in terms of simple model potentials).

The above question intrigued me for the first time when, based on the existing theory of poles of the S-matrix, eg., [69], on the result of Siegert [72], and on the previous work of Herzenberg, Mandl, et al [26], I wrote eq. (1) in [8], (eq. 3 here), as an *assertion*. At the time, I was unable to find a formally rigorous answer. This was achieved in collaboration with Komninos in 1980 [40], as outlined below. The same approach was later used for the analysis of the resonance states that are induced by an external field [12].

Fano [73] used the stationary superposition for prediagonalized basis functions with real energies,

$$\Psi(r,E) = a(E)\Psi_0(r) + \int_0^\infty b_E(E')\varphi(r,E')dE' \tag{7}$$

and formally solved for the energy-dependent coefficients, $a(E)$ and $b_E(E')$.

As pointed out in [8], of key importance in the theory of resonances is the asymptotic boundary condition in the description of the relevant wavefunctions. Accordingly, in [40,12] we started by recognizing that eq. (7) is a *standing wave*, and showed how, for large $r$, it is reduced to an equivalent solution in terms of the sum of two adjoint complex *traveling waves*, in the form,

$$\Psi(r,E) \xrightarrow{r \to \infty} -\sqrt{\frac{\pi}{2k}}Va(E)\left[\left(1-\frac{\lambda(E)}{i\pi}\right)e^{iN} + \left(1+\frac{\lambda(E)}{i\pi}\right)e^{-iN}\right] \tag{8}$$

where $V$ is the matrix element $<\Psi_0 \,|\, \mathbf{H} \,|\, \varphi(E)>$, and the phase $N$ corresponds to the different result for each potential. The value of $\lambda(E)$ is obtained by imposing the outgoing-wave boundary condition that a decaying state must have. Using eq. (8), this means that on resonance, the coefficient of the incoming wave must be zero, i.e., $\lambda(E) = -i\pi$. Therefore, using Fano's expression for $\lambda(E) =$ $(E-E_0-\Delta(E))/|V(E)|^2$, both the complex eigenvalue on resonance, and the asymptotic part of the eigenfunction emerge naturally, expressed in terms of computable quantities:

$$E_r = E_0 + \Delta(E_r) - i\Gamma(E_r)/2, \qquad \Psi_r^{as} \sim \sqrt{\frac{2\pi}{k_r}}Va(E_r)e^{iN} \tag{9}$$

So, it is the sum of $\Psi_r^{as}$ with $\Psi_0$ that produces the two-parts form of the complex eigenfunction on resonance, eq. (5) [5,40,12]. This form provides a guiding recipe for calculations, regardless of formalism and computational details. Apart from this, the asymptotic expression of $\Psi_r^{as}$ has been used for the derivation of a condition on resonance which has been applied for the optimization of "small" trial wavefunctions, e.g., [79].

The derivation of the many-electron CESE outlined above is rigorous, and justifies eq. (3) which was stated as an assertion in [8]. The remaining problem has to do with the correct and consistent computation of matrix elements using resonance wavefunctions, since they are not square-integrable. A practical method is the use of complex coordinates, introduced in 1961 for short-range potentials (nuclear physics) by Dykhne and Chaplik [93]. A related discussion can be found in [5].

According to the foregoing discussion, and to results regarding the invariance of $<\Psi_0 \mid \mathbf{H} \mid \Psi_0>$ under coordinate rotations in $\mathbf{H}$ and in $\Psi_0$ [39], the SPS form of trial wavefunctions for the non-perturbative calculation of resonance complex eigenvalues (CESE method) is,

$$\tilde{\Psi} = a_0 \Psi_0 + \sum_n a_n \tilde{u}_n \qquad \text{(eq. 7.7 of [39])} \qquad (10)$$

where $\tilde{u}_n$ are N-electron configurations for asymptotic correlation, containing (N-1)-electron terms with real coordinates coupled to complex functions for the outgoing electron in each open channel [5]. Unlike the case of the theory and computational implementation of the CCR method, in the CESE method the coordinates of $\mathbf{H}$ are real.

The choice of the function spaces representing $\Psi_0$ and $\tilde{u}_n$ depends on the problem. In our treatment of atomic resonances, $\Psi_0$ normally consists of both MCHF numerical and correlation analytic orbitals. When the MCHF solution is impossible, as in the case of the $H^-$ resonances [49-51], a systematic choice of diffuse functions replaces the numerical HF ones, reaching the asymptotic region.

### 5.1.4. Exterior complex scaling and analytic continuation involving the asymptotic region

The proposal discussed in [63], concerning the systematic calculation of matrix elements of $\mathbf{H}$ and of $\mathbf{H}^2$ involving resonance wavefunctions, later named the method of '*exterior complex scaling*', is in the same spirit, namely, that of recognizing in practice the different roles played by $\Psi_0$ and by the asymptotic component of the resonance eigenfunction.

Specifically, the ECS regularization procedure proposed in [63] was written as,

$$\int\limits_{all\ space} \psi^2 dr \ = \ \int\limits_0^R \psi^2 dr \ + \ \int\limits_C \psi^2 ds \qquad\qquad \text{(eq. 3 of [63])} \qquad\qquad (11)$$

where R is a point on the real axis at the edge of the inner region and $R < \operatorname{Re} s < \infty$.

I quote from [63]. '*Construction (3) shows that, as the resonance width tends to zero, the contour integral is brought to the real axis and the integral* $\int\limits_{all\ space}$ *is finite on the real line because it is taken over bound states. This suggests that the square-integrable function* $\Psi_0$ *and Gamow's [resonance] function can be thought of as being related via analytic continuation*'.

For early numerical applications to various potentials, including that of the 'volcanic' ground state of $He_2^{++}$, see [94].

## 5.2. Topic II

The solution of MEPs for this topic has been achieved via the construction and implementation of the *many-electron, many-photon theory*, which is a non-perturbative, non-Hermitian, SPS formulation in the spirit of the CESE theory for field-free resonances. Lack of space does not permit elaboration. The reader is referred to section 11 of [5] and to [9-14, 95]. Hamiltonians with electric (static and periodic) and magnetic static fields have been used in prototypical applications. For example:

1) Energy shifts and single-electron ionization transition rates for absorption of one or more photons have been calculated, for one, two and three-color fields, e.g., [9-14,5].

2) Quantitative studies were carried out on doubly excited states perturbed by strong fields, whereby many continua are mixed, and positions and widths change due to the perturbation. E.g., see [13] for DESs of $H^-$ and $He$ in ac and dc electric fields, and [95] for DESs of $H^-$ in a static magnetic field.

3) Cross-sections for the photo-ejection of two-electrons from the closed-shell $1s^2\ {}^1S$ state of $He$ and $H^-$, as well as from the open-shell metastable state $1s2s2p$ ${}^4P^o$ of $He^-$, were determined for energies very close to threshold [96,97].

### 5.2.1. Field-induced resonances from analytic continuation of large-order perturbation series

A formally possible alternative to nonperturbative variational methods for the calculation of complex energies of resonances is the construction and implementation of methods of *large-order perturbation theory* (LOPT) in conjunction with techniques of summation. The subject of LOPT is rather esoteric. Even though such approaches have features of mathematical elegance, the reality of the MEP has restricted their application to one-electron systems. Reviews of LOPT methods as applied to resonances created by the application of electric fields to Hydrogen were given by Silverstone [98] and by Silverman and Nicolaides [99] in the book [100].

In subsection 5.1.4, I quoted a statement from [63] regarding the possibility of direct connection between the square-integrable $\Psi_0$ and the resonance eigenfunction, $\Psi_r$, via the change of boundary conditions and analytic continuation.

Within a different framework, Reinhardt [101] produced an enlightening analysis and justification regarding analytic continuation for the calculation of the ground state resonance of Hydrogen in a strong electric field in terms of Padé summation of the divergent LOPT Rayleigh-Schrödinger series.

In the approach of Silverman and Nicolaides [99,102,103], complex energies were calculated to high accuracy not only for the ground state, but also for hundreds of degenerate excited levels of Hydrogen. The analytic continuation is accomplished by shifting the origin of the real eigenvalue series into the complex plane, where the relevant divergent series are summed by a twofold application of Padé approximants. Because this formalism is based on expansions over state-specific function spaces, the formalism is applicable to N-electron problems.

### 5.3. Topic III

The solution of the METDSE is achievable in terms of the SSEA [4,6, 15-17]. The SSEA is conceptually simple, as well as general. For each problem of interest, we first make an approximate analysis of the apparent requirements set by the pulse characteristics and the spectrum of the stationary states. Then, the SSEA solution, $|\Psi(t)>_{SSEA}$, is constructed and computed in the form, (I omit the index for each possible channel),

$$|\Psi(t)>_{SSEA} = \sum_m a_m(t)|m> + \int_0^\infty b_\varepsilon(t)|\varepsilon>d\varepsilon \qquad (12)$$

The expansion (12) holds for atoms as well as for molecules. $|m>$ represents the relevant discrete *state-specific* wavefunctions and $|\varepsilon>$ are energy-normalized state-specific scattering states. Resonance states are represented by their localized component. The mixing complex coefficients are time-dependent. The Hamiltonian is either $\mathbf{H}_{A(M)}$ for autoionization, with $\Psi_0$ being the initial ($t=0$) wavepacket, or the full $\boldsymbol{H}(t)$ of eq. (2) for atom (molecule) – pulse interaction. With these Hamiltonians, and with the state-specific wavefunctions constituting $|\Psi(t)>_{SSEA}$, numerically accurate bound-bound, bound-free and free-free matrix elements are calculated.

Substitution of (12) into the METDSE produces a system of coupled equations containing energies, matrix elements, and the unknown coefficients. In the process of its solution, one can monitor and evaluate with transparency and economy the dependence of the evolution of $\Psi(t)$ of eq. (2) on each $|m>$ and $|\varepsilon>$. Since these stationary wavefunctions are state-specific, the solution of the METDSE according to the SSEA takes into account,

- The zero-order features of electronic structures of initial, intermediate and final states. These can be calculated at the HF or MCHF levels.
- The dominant electron correlations, for those states where this is necessary.
- The presence of perturbed or unperturbed Rydberg levels and multiply excited states.
- The contribution of the continuous spectrum of energy-normalized, channel-dependent scattering states, without or with resonance states.

- The interchannel coupling.

See [4,6] and their references.

A rather spectacular spectroscopic result in this direction has been the experimental and theoretical discovery that, at the fundamental level of electronic structure and dynamics of atoms and molecules, there is a *relative time delay* in the emission of different electrons upon absorption of an energetic photon [104]. The theory and many-electron approach used in [104] for the treatment of the dynamics of the photoejection of the $2s$ and $2p$ electrons of Neon by an attosecond pulse, are explained again in [4].

## 6. EPILOGUE

Topics I, II and III of section 2, with which this paper has dealt, belong to the high-lying portion of the "excitation axis" of **Figure 1** and contain broad categories of *resonance/non-stationary* states and of electronic dynamics involving the continuous spectrum near and far from threshold, for Hamiltonians without and with external strong fields.

After a brief review of the situation that existed with theory and calculations on N-electron systems when our approaches were introduced, (section 4), I outlined how the requirements for the solution of a variety of *many-electron problems* have been satisfied in terms of efficient *state- and property-specific* (SPS) formalisms and computational methodologies. These have been developed and implemented within generally applicable single- and multi-state Hermitian (real energies) and non-Hermitian (complex energies), time-dependent and time-independent frameworks, which place emphasis on the proper *forms* of the trial wavefunctions, as in eqs. 5, 10 and 12, and on the choice and use of corresponding function spaces.

The methods of calculation of correlated wavefunctions beyond the zero-order HF or MCHF levels are *variational* and constitute the standard SPS approach. In addition, in the case of time-independent formulations of problems of field-induced resonances, it has been demonstrated that the implementation of *large-order perturbation theory* is also possible, using a formalism which incorporates analytic continuation and efficient summation of divergent series, and is in principle applicable to many-electron ground and excited states (see subsection 5.2.1).

The core ideas of the SPS theory are stated in section 3. Reviews of the SPS methodology and of applications to the topics discussed here, can be found in [3-7].

The corresponding results for wavefunctions, properties and phenomena are numerous, covering a broad spectrum of prototypical cases. Where experimental data are available, the comparison between them and those obtained from SPS calculations shows very good agreement.

**ACKNOWLEDGMENTS:** I am thankful to Yannis Komninos and Theodoros Mercouris for our decades-old collaboration on the topics discussed in this paper.